**A simulation-based evaluation of a Cargo-Hitching service for E-commerce using mobility-on-demand vehicles.**

André Alho[a]* (0000-0003-3275-7174)
Takanori Sakai[a] (0000-0002-8426-7452)
Simon Oh[a] (0000-0001-7013-9997)
Cheng Cheng[a] (0000-0002-4567-4098)
Ravi Seshadri[a] (0000-0002-9327-9455)
Wen Han Chong[a] (-)
Yusuke Hara[a] (0000-0003-0268-9908)
Julia Caravias[a] (-)
Lynette Cheah[b] (0000-0001-6312-0331)
Moshe Ben-Akiva[c] (0000-0002-9635-9987)

[a]*Singapore-MIT Alliance for Research and Technology, 1 CREATE Way, #09-02 CREATE Tower, Singapore 138602*
[b]*Engineering Systems and Design, Singapore University of Technology and Design, Singapore*
[c]*Intelligent Transportation Systems Lab, Massachusetts Institute of Technology, Cambridge, M.A., 02139, U.S.A.*

*Corresponding author
Postal address: 1 CREATE Way, #09-02 CREATE Tower, 049374, Singapore
E-mail: andre.romano@smart.mit.edu
Telephone: +65-6601-1636

**ABSTRACT**
Time-sensitive parcel deliveries, shipments requested for delivery in a day or less, are an increasingly important research subject. It is challenging to deal with these deliveries from a carrier perspective since it entails additional planning constraints, preventing an efficient consolidation of deliveries which is possible when demand is well known in advance. Furthermore, such time-sensitive deliveries are requested to a wider spatial scope than retail centers, including homes and offices. Therefore, an increase in such deliveries is considered to exacerbate negative externalities such as congestion and emissions. One of the solutions is to leverage spare capacity in passenger transport modes. This concept is often denominated as *cargo-hitching*. While there are various possible system designs, it is crucial that such solution does not deteriorate the quality of service of passenger trips. This research aims to evaluate the use of Mobility-On-Demand services to perform same-day parcel deliveries. For this purpose, we use SimMobility, a high-resolution agent-based simulation platform of passenger and freight flows, applied in Singapore. E-commerce demand carrier data are used to characterize simulated parcel delivery demand. Operational scenarios that aim to minimize the adverse effect of fulfilling deliveries with Mobility-On-Demand vehicles on Mobility-On-Demand passenger flows (fulfillment, wait and travel times) are explored. Results indicate that the Mobility-On-Demand services have potential to fulfill a considerable amount of parcel deliveries and decrease freight vehicle traffic and total vehicle-kilometers-travelled without compromising the quality of Mobility-On-Demand for passenger travel.

# 1. INTRODUCTION

The rapid growth of e-commerce deliveries over the last decade is bringing about changes to the freight transportation sector. E-commerce retail sales in the U.S. represented 4% of the total retail sales in 2010 rising to 11.8% in 2020 (U.S. Department of Commerce, 2020) and the rate of e-commerce adoption increased during the COVID-19 pandemic (Adobe Analytics, 2020) with a potential lasting effect. An increase in e-commerce-derived shipments leads to a "fragmentation" of shipment sizes, since these e-commerce-derived shipments are smaller and spatially/temporally more spread than conventional Business-to-Business (B2B) shipments such as those to warehouses and retail stores (Morganti et al., 2014). Moreover, express services for e-commerce deliveries (Dablanc et al., 2017) and high return rates contribute to a process inefficiency and can lead to externalities such as congestion and emissions. For example, in Germany, 77% of shoppers have returned products and one third of all distributors have average return rates of over 20% partially due to incentives such as "free-of-charge" returns (Morganti et al., 2014). It can be hypothesized that e-commerce deliveries to consumers have room to improve their efficiency.

Transportation modes that serve trips performed by individuals often have some spare capacities, which can be used for freight, such as parcels. This concept is termed cargo-hitching (Trentini et al., 2010; Sampaio et al., 2017). The joint use of passenger airplanes to carry freight is an example of cargo-hitching. However, there is limited research exploring cargo-hitching in urban settings. Van Duin et al. (2019) report on a series of urban cargo-hitching initiatives across Europe and in India. While the authors highlight the promising benefit of cargo-hitching, they also point out the uncertainty about its long-term viability attributable to a lack of an understanding on viable business models. Crowd-logistics is a type of cargo-hitching scheme (Rai et al., 2017). Its crux is the use of occasional carriers with spare capacity to perform some task (e.g. freight pickups/deliveries) and, ideally, it should leverage vehicle trips which are similar to the shipment origin/destination, while not compromising the service quality for passenger travel. Service quality can be defined by attributes such as securing a ride and minimizing waiting and travel times.

Recent real-world applications of cargo-hitching came to the spotlight during the COVID-19 pandemic, where Mobility-On-Demand (MOD) services - Transportation Network Companies and conventional taxis - were used to increase the capacity of logistics services to support grocery deliveries (Kai Yi, 2020, Ho, 2020). We argue that the MOD system is comparatively more suitable to handle time-sensitive deliveries - requested for delivery in a day or less - since these would have to fit into currently planned tours by carriers, potentially leading to less efficient operations. Still, there are some questions that we believe particularly relevant for the viability concerns raised by Van Duin et al. (2019), namely how much freight can be moved and what are the potential impacts in the level of service for passengers under a non-extreme scenario where passenger travel is still occurring.

In this paper, we study the application of cargo-hitching to mobility-on-demand (MOD) vehicles, exploring how operational assumptions – i.e., how shipments are assigned to MOD trips – influences the resulting flows, namely in changes to passenger travel and in reductions of dedicated freight-vehicle flows.

It must be noted that cargo-hitching in an application to MOD is not suitable for all shipment types. *Suitability* can be defined by the characteristics of shipments such as *volume/weight* and *commodity type* (i.e., shipments which are large and/or require special equipment cannot be handled by non-professional freight carriers), *transaction type* (i.e., the volumes of B2B transactions might not be suitable for cargo-hitching), and *temporal nature (i.e.,* if a conventional carrier knows the shipment demand in advance, it can plan for deliveries in an efficient manner). It should also be noted that the acceptance of a delivery task is subject to specific factors relevant to the transporting and the receiver parties (Paloheimo et al., 2016; Devari et al., 2017; Punel and Stathopoulos, 2017, 2018; Ermagun and Stathopoulos, 2018). Thus, the case can be made for a specific type of shipments - parcels - which are typically small, lightweight, and can be carried without special equipment, as well as being the *de facto* shipment unit for e-commerce deliveries. We use the term *parcel* to refer to shipments suitable for cargo-hitching services and focus on such shipments.

This paper contributes to the existing literature on cargo-hitching in the following dimensions: 1) Applying an agent-based simulation framework to systematically investigate the impacts of cargo-hitching from the perspective of travelers, carriers, and regulators. The simulation framework includes the detailed modelling of mobility-on-demand services on the demand/supply sides, explicitly capturing demand-supply interactions. 2) Performing extensive simulations to cover different assignment strategies of freight demand to MOD vehicles using a city-wide model of Singapore in 2030 and yielding insights into the potential impacts of cargo-hitching.

## 2. LITERATURE REVIEW

This literature review mainly focusses on applications of cargo-hitching to taxi and mobility-on-demand. There are other application domains for cargo-hitching which are not covered in detail such as applications involving users of public transport (Shen et al., 2015; Pimentel and Alvelos, 2018; Serafini et al., 2018; Gatta et al., 2019) or private cars (Devari et al., 2017; Arslan et al., 2018). There is also a large body of research on participation to such initiatives from the perspective of occasional carriers (Paloheimo et al., 2016, Devari et al., 2017; Punel and Stathopoulos, 2017, 2018; Ermagun and Stathopoulos, 2018), and of professional carriers (Huang and Ardiansyah, 2019), both being out-of-scope of this study. Lastly, we focus on business-to-consumer flows, despite the existence of research on reverse freight flows, such as returned parcels (Chen et al., 2016).

A common thread across most experiments involving cargo-hitching and modes such as taxi or mobility-on-demand is that the motivation of the research is a proposal of a matching algorithm. We can define matching algorithms as formulation to solve assignment problems by pairing demand (requests for a ride) and supply (vehicles), in various application settings (e.g. online or offline). To demonstrate their potential, the authors typically follow the description of the algorithm with an application to illustrate its performance and derive some insights.

Li et al. (2014) proposed matching algorithms for two cases, the first where passengers and parcels are considered simultaneously, and the second where freight is inserted into pre-computed passenger travel routes. The authors use synthetic freight data in a San Francisco case-study. The results indicate the influence of the spatial distributions of requests on the fleet performance (and thus profitability), with sharing strategies being fitter for denser urban areas. Nguyen et al. (2015) demonstrate benefits of sharing by applying two matching algorithms for deploying taxies serving passengers and parcels with the objective of finding valid service options that achieve the maximum profit. The authors consider a static case where all demand is known in advance under a direct demand-serving strategy and a shared-ride strategy. The study shows that, even when a match rate is small (~8%) in a shared-ride strategy, there are some savings compared to direct-demand service across a range of indicators such as cost, travel distance and the number of taxis required to move parcels. However, no comparison with the case in which freight is served by conventional vehicles is provided. Chen et al. (2017) present a method to explore the potential for chained taxi trips to move parcels, with stopovers at package interchange stations, which are 24-hours convenience stores. Their case study uses synthetic freight demand in Hangzhou, China. Taxis are assigned exclusively to parcel movements during off-peak hours and before passenger pickup and after passenger drop-off. Furthermore, the authors mention required incentives for taxi drivers to participate in the initiative, which might limit the practical feasibility of this proposed solution. Najafabadi and Allahviranloo (2019) develop an algorithm to match taxis and delivery requests in a real-time dynamic setting, aiming to minimize distance travelled, transportation cost and the number of vehicles used. They use parcel delivery demand estimated based on a household travel survey. Some benefits are identified for some scenarios; however, in these scenarios, passenger travel times are increased by 8% to 13% but have inversely proportional cost savings due to the sharing setting.

Neither of the studies above investigate thoroughly the implication of the assumed sharing settings or conduct a comparison with freight demand and associated freight traffic demand without cargo-hitching services. Regarding the first, we hypothesize the assumed assignment process of parcels to the cargo-hitching vehicles has a significant influence on the outcome of the analysis. For the latter, the comparison can be justifiable if demand and supply models reasonably replicating spatial-temporal characteristics of passenger and freight demand. Namely, parcels are typically transported in multi-stop delivery tours rather than in dedicated trips, which can influence (negatively) the relative performance of freight vehicles.

In summary, most existing literature focuses on the performance and scope of matching algorithms, demonstrating them using simulation-based methods. We identify room for improvement with regards to the exploration of assignment strategies, i.e. the assumptions behind the assignment of freight to passenger trips, and an adequate representation of passenger and freight flows, i.e. using state-of-the-art models representing accurately relevant passenger and freight flows. This research aims to derive the insights from experiments, rather than the methodology behind the matching algorithm or demand/supply models, which are subject matter of prior published research. In the following section, we detail an application of an agent-based simulator to the city-state of Singapore, which is used for such experiments.

## 3. SIMULATION FRAMEWORK

The evaluation of the network-wide impacts of cargo-hitching requires a high-fidelity simulation framework that models freight and passenger demand, supply, and their interactions. We use SimMobility, the agent- and activity-based simulation laboratory, - which integrates disaggregate behavioral models at multiple time scales in a consistent and coherent manner and simulates both urban freight and passenger travel (see Alho et al., 2020; Adnan et al., 2016 for more detailed description). The overall framework of the SimMobility platform is shown in Figure 1, which consists of freight and passenger models organized into three modules: Long-term, Mid-term and Short-term.

The Long-term module captures land use and economic activity, with an emphasis on accessibility. It predicts the evolution of land use as well as property development and use, determines the associated life cycle decisions of agents, and accounts for interactions among individuals (Zhu et al., 2018; Zhu and Ferreira Jr, 2014) and businesses (i.e., commodity flows as in Sakai et al., 2020a) as well as overnight parking choices of freight vehicles (Gopalakrishnan et al., 2020). On the passenger side, the Mid-term models address daily transportation demand and supply for passenger travel; it simulates agents' behavior including their activity and travel patterns and the movement of vehicles at a mesoscopic level (mesoscopic traffic simulation). On the freight side, the Mid-term simulates "Pre-day" logistics planning, "Within-day" freight vehicle operations, and traffic movements (mesoscopic traffic simulation) (Sakai et al., 2020b). The Short-term simulator is a multimodal microscopic traffic simulator that functions at the operational level; it simulates the movement of agents at a granularity of milliseconds, and synthesizes driving and travel behavior models, traffic control and management systems, and the simulation of communications networks (Azevedo et al., 2017).

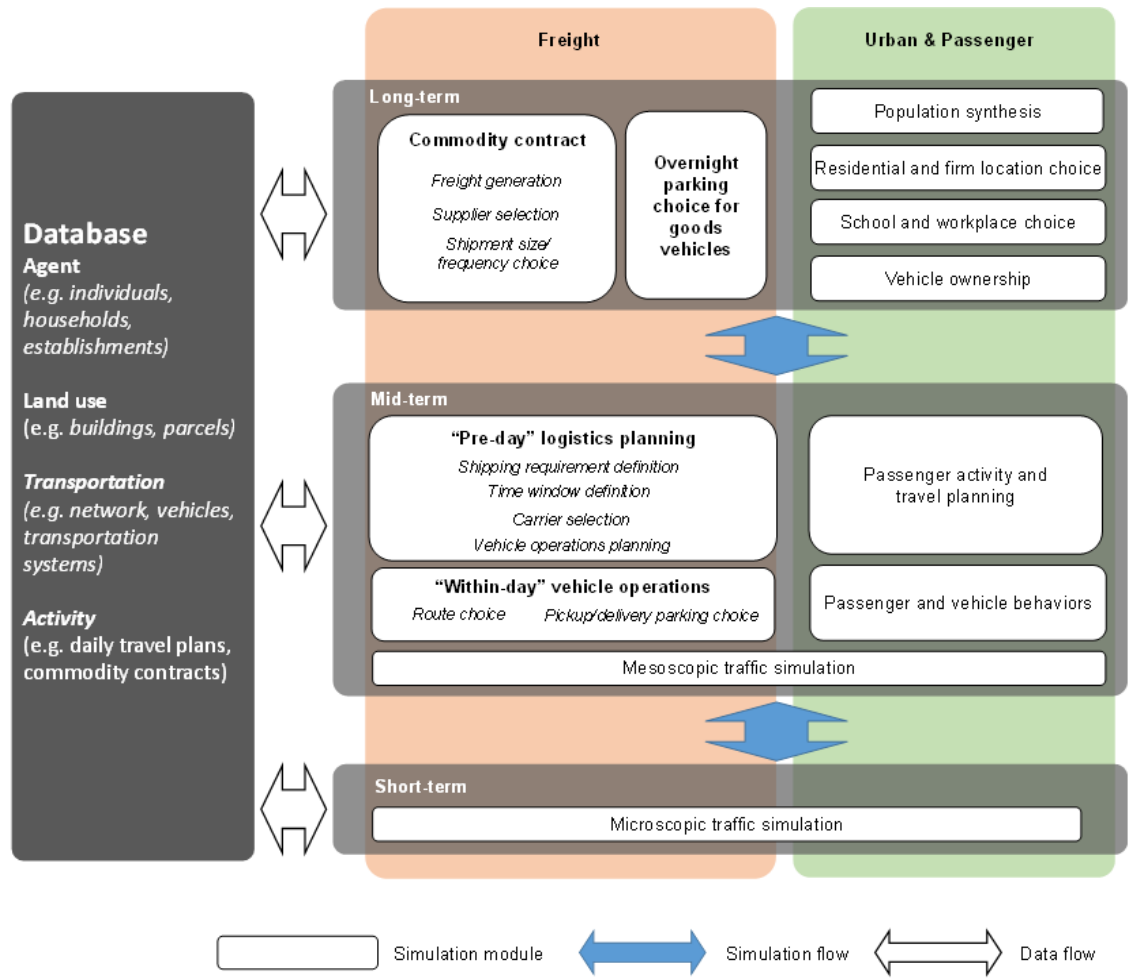

**Figure 1. Framework of SimMobility (Modified from Sakai et al., 2020b).**

In this paper, we primarily use the Mid-term module, which consists of three sub-modules, the Pre-day (passenger and freight), Within-day (passenger and freight) and Supply. Together, these modules simulate daily passenger and freight demand, supply, and demand-supply interactions.

The Pre-day module generates travel demand in the form of daily activity schedules for each individual in the population, and freight activity schedules for each vehicle in the fleet of carriers in the establishment population. On the passenger side, the Pre-day module is an activity-based model (ABM) system that adopts the Daily Activity Schedule approach (Ben-Akiva et al., 1996) to predict agents' activity-travel schedule (plan) including: (1) activity types and number, (2) activity durations and time-of-day, (3) activity locations, and (4) modes. The Pre-day ABM is a system of hierarchical discrete choice models (logit and nested-logit) organized into three levels (day pattern, tour and stop level), and takes as an input a detailed synthetic population of individuals and households including sociodemographic characteristics and vehicle ownership (refer to Oh et al., 2020a for more details).

On the freight side, the Pre-day module (Sakai et al. 2020b) handles the conversion of shipment demand to vehicle tours. It is a system of models that simulates shipping requirement definition, time-window selection, carrier selection and vehicle operation planning (shipment-to-vehicle assignment and tour formation). These models take as inputs a list of commodity contracts (i.e., selling and purchasing policies), which include the information of commodity type, shipper-receiver pairs, and delivery frequency and weight and provides a shipment list to be handled in an average day. The output - vehicle operations plans (VOPs) - include details such as planned stop locations, arrival, and departure times, and stop purposes (overnight parking, pickup or delivery). These plans, essentially outlining vehicle tours, are used as the inputs for a mesoscopic or microscopic traffic simulation. The demand models were initially built to deal with Business-To-Business (B2B) demand, and, for this paper, a scenario of Business-To-Consumer (B2C) demand is assumed and further described in Sections 4.1 and 4.2. These B2C shipments - parcels - are assumed to be exclusively transported with by businesses defined as *parcel carriers*. Henceforth, B2C shipments will be referred to as parcels.

The Within-day module, on the passenger side, includes models of departure time choice, route choice and within-day rescheduling, which converts plans into actions based on the plan-action framework (Ben-Akiva, 2010). Freight models include dynamic route choice and en-route parking choice. Finally, the Supply module takes the detailed trip chains (from the Pre-day and Within-day) and simulates the movements of freight and passenger vehicles as well as individuals on a multimodal network. The Supply module consists of a mesoscopic traffic simulator integrated with various 'controllers' that model the operations of fleet operators, including public transit (bus and train controllers) and on-demand services (smart mobility service controller). The bus and rail controllers operate the scheduling of vehicles (frequency/headway based), monitor the vehicle occupancy, and determine dwell time of the vehicle at stops/stations. The smart mobility service controller (called hereafter SMS controller) simulates all aspects of the operations of an on-demand service including receiving ride-requests for single or shared taxi rides, assignment of vehicles to requests, dispatching and routing of vehicles, and rebalancing of vehicles when idle.

More specifically, passengers send ride requests in the form of an origin and destination (pick up and drop off locations) to the controller. The SMS controller periodically processes these requests and performs vehicle assignments (from the vehicle fleet) to the individual requests. Additional actions of the controller include rebalancing idle vehicles (for example, to zones of higher demand) or directing these vehicles to suitable parking or holding locations. The actions of the controller are communicated to individual vehicles in the form of modifications to a schedule, which is maintained for each vehicle in the fleet at any point in time and consists of a sequence of schedule items. The schedule item includes specific actions such as pick-up, drop-off, cruise, park, and associated information such as time and locations for pick-ups, drop-offs, cruising, and parking.

The assignment of requests to vehicles is performed using a simple insertion heuristic. To maximize the extent of ride sharing, the controller attempts to first match shared-ride requests to vehicles that are already serving one or more requests. If this is not possible, it attempts to match the request to idle vehicles. The process of matching a request to a vehicle involves iterating through the fleet of vehicles and finding the first vehicle that can satisfy the incoming request whilst meeting certain constraints on waiting time and travel time of both the incoming request and existing requests in the schedule of the vehicle. Thus, for each vehicle, candidate schedules are created by inserting the pick-up and drop-off items of the incoming request within the existing schedule. If a candidate schedule is found that meets waiting time and travel time constraints of the incoming passenger and existing passenger(s), the schedule is deemed feasible and the request is assigned to the vehicle. More details on the controller may be found in the past research (Oh et al., 2020a, Nahmias-Biran et al., 2019, and Basu et al. 2018). Additional assumptions relating to the controller and its adaptation to model cargo-hitching are discussed in Section 4.2.

Finally, demand-supply interactions are explicitly modeled through two iterative learning mechanisms, day-to-day and within-day learning, which involve performing several iterations (or runs) of the Pre-day, Within-day and Supply simulations to achieve consistency (or equilibrium) between demand and supply. More specifically, the within-day learning component involves an iterative update of time-dependent link travel times and public transit waiting times resulting in an adjustment of route choice decisions until an equilibrium is attained (equilibrium here refers to the fact that the route travel times perceived by users in making their route choice decisions and the actual or realized' travel times are within a pre-specified tolerance). The day-to-day learning component involves an iterative update of aggregate zone to zone travel times and waiting times (of both public transit and on-demand services) resulting in an iterative adjustment of travel and activity decisions of individuals until equilibrium or constancy is achieved. Equilibrium here refers to the fact that the zone to zone travel times and waiting times perceived by users in making their activity and travel decisions and the actual or realized' travel times/waiting times are within a pre-specified tolerance.

## 4. EXPERIMENT DESIGN

This section details the experimental design, including the SimMobility application to Singapore, detailing the setup of simulations (4.1), the scenarios (4.2), and the metrics to quantify system changes (4.3).

### 4.1 SimMobility Application to Singapore

The simulations of Cargo-hitching scenarios are conducted using a SimMobility model of Singapore in 2030. The synthetic population for 2030 (6.7 million individuals) was generated using a Bayesian approach (Sun and Erath, 2015) based on socio-economic data, land-use data, and relevant control totals (see also Zhu and Ferreira, 2014 for more details on the population synthesis). The demand model for 2030 relies on a calibrated activity-based model system for the year 2012 (that matches observed tour/stop generation rates, activity shares, and modes closely) estimated using Household Travel Survey data (Oh et al., 2020a; Lu et al., 2015; Li, 2015). This model was enhanced with the MOD modes, by assuming a similar utility specification as taxis and calibrating relevant alternative specific constants (ASCs) and scale parameters of the mode and mode-destination choice models against aggregate data on the usage of single ride and shared MOD modes (see also Nahmias-Biran et al., 2019 for more details). The calibration and validation also included matching simulated outputs to observed screen-line counts, public transit smart card data and network travel times (for more details on the model calibration, the reader is referred to Oh et al., 2020a). The freight model is calibrated against 2012 screen-line traffic count data, applied to a predicted 2030 business establishments which considers land-use plans and employment growth as well as available vehicle fleets. Parcel demand is separately considered as per the scenarios described in section 4.1.2. The road and transit networks of the year 2030 in Singapore consist of 1,169 zones, 6,375 nodes, 15,128 links, 730 bus lines covering 4,813 bus stops, 26 MRT lines over 186 stations. The road network is shown in Figure 2.

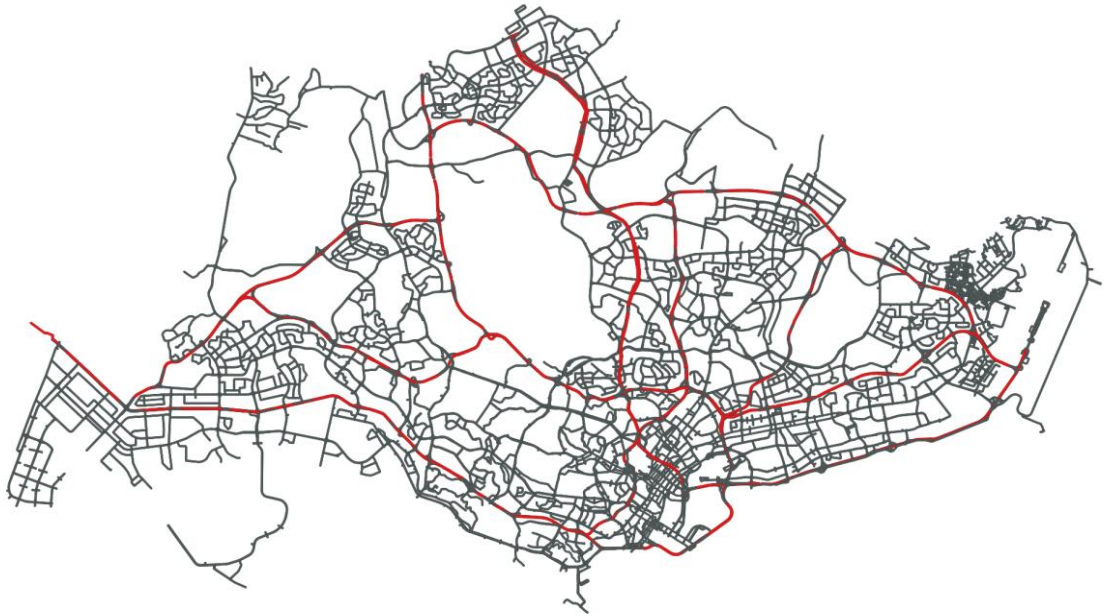

**Figure 2. Singapore Network (red: highway, grey: arterial)**

*4.1.1 Demand for MOD*

The characteristics of the demand for MOD services and the e-commerce delivery demand, which are central to the scenario simulations, are as follows. The passenger demand for MOD services from the Pre-day module is 576,786 trips for a 24-hour period (a mode share of 6%). The temporal distribution of MOD demand is shown in Figure 3. It shows a typical commuting pattern with demand surges in the AM and PM peak periods. Requests for shared rides represent 27% of total MOD requests. The average travel times for single and shared rides are 13 minutes for single and 17 minutes for shared rides over, with average distances travelled of, respectively, 12 km and 15 km.

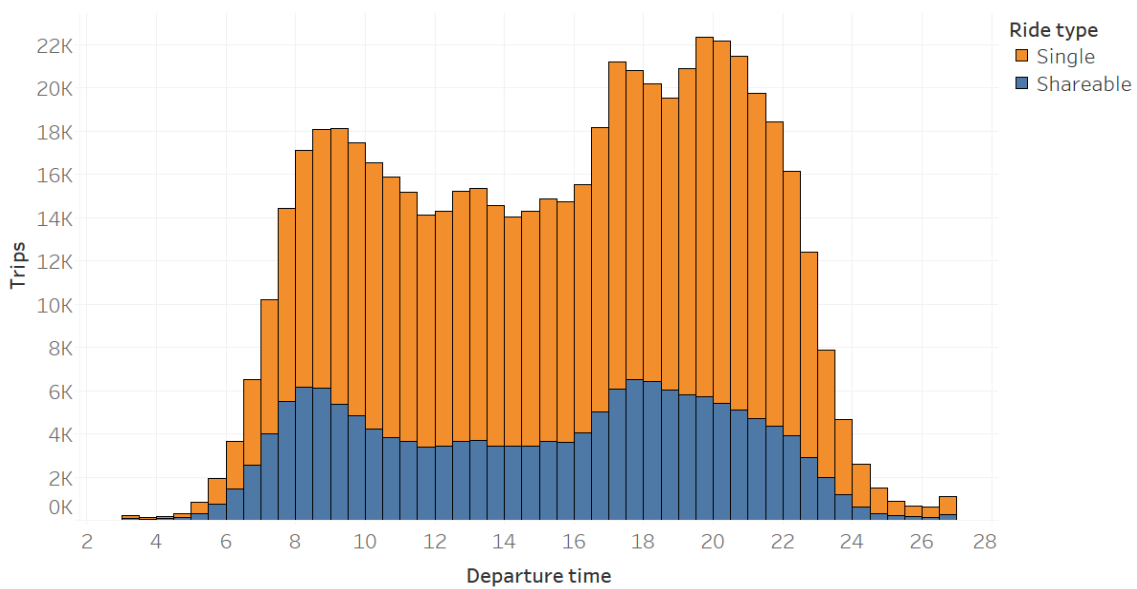

**Figure 3. Passenger demand for MOD**

*4.1.2 B2B and B2C (parcels) shipments*

We assume parcel demand predominantly e-commerce-derived and generate a set of same-day parcel deliveries which are added to our base demand already considering B2B movements. We assume a scenario of 67,000 same-day deliveries requested to the MOD operator, which represents around 34% of total present-day daily e-commerce delivery demand (Wong, 2020), and around 12% of total MOD requests, and 43% of shared MOD requests in the simulation setting. This experiment does not explicitly model other types of B2C deliveries which are considered a further research subject. We acknowledge those would add to the total demand handled by the freight vehicles but nonetheless remain constant across scenarios.

For purposes of the simulation, we randomly sample delivery requests from e-commerce delivery records of a carrier in Singapore. The e-commerce delivery records detail parcel origins, destinations, and successful delivery time. The information about delivery time-windows is lacking in the records. We do not consider the time-windows in this analysis while assume the delivery time as the request time. The temporal distribution of parcel requests to the MOD vehicles is shown in Figure 4.

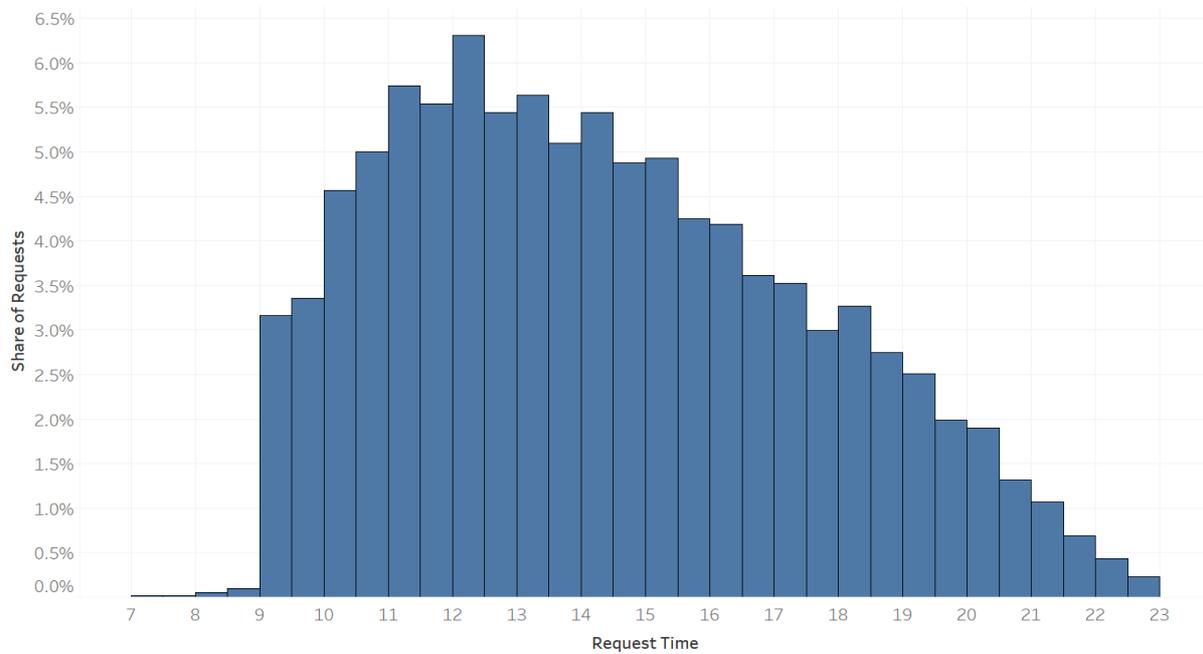

**Figure 4. Temporal distribution of freight demand for B-to-C same-day parcels delivery during a day**

An illustration of the resulting vehicle trips for predicted B2B and B2C "same-day" shipments is shown in Figure 5. The total number of trips is 1.18 million.

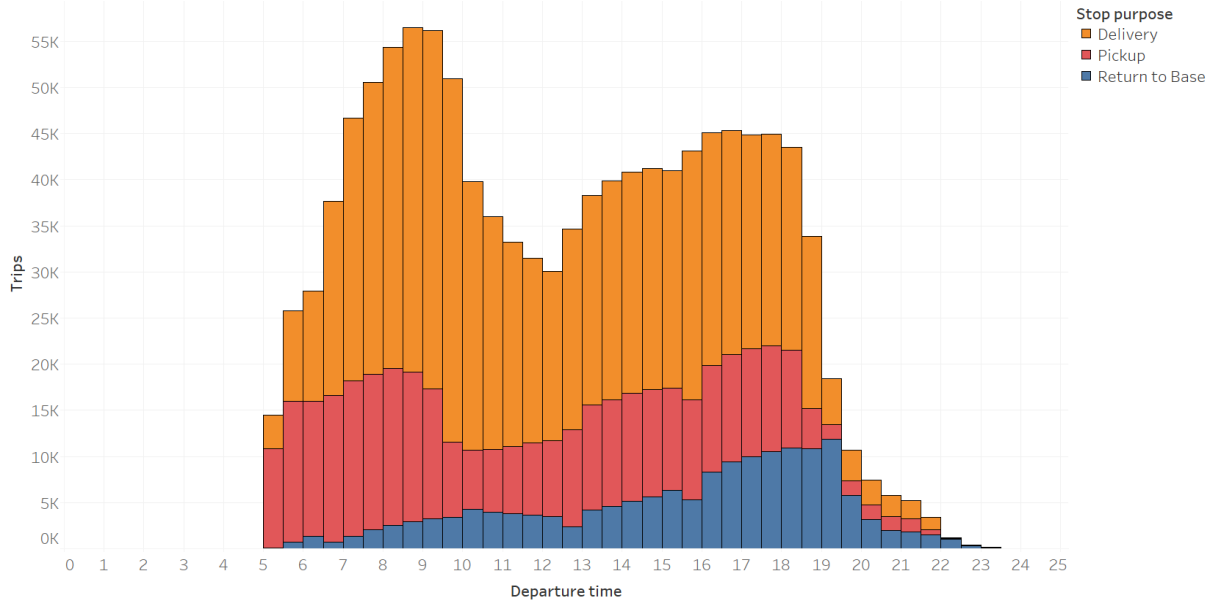

**Figure 5. Freight vehicle trips by purpose inclusive of B2B and B2C "same-day" shipments**

### 4.2 Scenarios

We define the cargo-hitching scenarios that involve the MOD fleet serving passenger and e-commerce parcel deliveries and vary as for operational configurations of the MOD service. In all scenarios, the MOD fleet size is assumed to be equal to 17,900. The fleet size is set to ensure a high request satisfaction rate, a high vehicle utilization during the peak period, and reasonable waiting times, based on the iterative simulations. Specifically, 96% of total passenger MOD demand (described in Section 4.1.1) is served, with 48% of requests being served in 5 minutes or less, and 85% of requests served within 10 minutes. For a comparison with existing MOD fleet sizes in Singapore and as of 2019, the total of equivalent vehicles (termed "private hire") is 45,000 (Channel News Asia, 2019); these vehicles are not operating 24 hours a day, but rather according to the availability of the drivers, with several drivers only working for a few hours a day (e.g., as a secondary job). In our scenarios, vehicles are assumed available for 24 hours a day. Thus, our fleet size assumption is a plausible representation of present-day fleets.

The following settings are made for simplifying the simulations of MOD fleet:
- Vehicle assignment does not consider delivery/ride fees. In other words, real-time effects of spatial demand and supply imbalances on ride prices (i.e., dynamic pricing or surge pricing are not modeled) and the differences in delivery and passenger travel fees are not considered.
- The assignment follows an online approach (i.e., real-time), and demand is not known *a-priori* to the controller; thus, the controller does not reposition vehicles considering expected or anticipated demand. Vehicles are initialized at random locations across the network.
- The controller handles both non-shared and shared ride requests, and the choice of a single or shared ride (made by the individual user) is determined by the demand models. Each ride request is assumed for a single passenger.
- A maximum waiting time threshold is 10 minutes for requests by passengers (the waiting time constraint within the insertion heuristic is described earlier in Section 3). The requests that remain in the booking system for longer than this threshold are considered "failed" and these passengers are assumed to use public transit.
- Dwell time is same for passenger and parcel pickups/drop-offs.
- A parcel takes the space equivalent to that of a passenger seat.

The operational scenarios are defined as follows:
- Baseline (Base): The MOD service (SMS controller) only serves passenger trip requests for single/shared ride services. All parcels are handled by conventional carriers.
- Shared (SHR): Parcel delivery requests to the SMS controller are considered for shared rides but a passenger must already have been assigned to the vehicle for a parcel request to be accepted. Additional requests up to the fulfilment of vehicle capacity can be accepted at any point of the ride subject to the constraints of the new passenger (waiting time) as well as the detour to the passengers(s) in the vehicle (tolerated delay) and those already scheduled for pickup (waiting time and tolerated delay).
- Shared and Idle (SHR+IDL): This is a variant of SHR where if no match to a shareable passenger ride can be found, and if idle vehicles are suitable to accept the parcel delivery request, these will carry the parcel. In more detail, for a given assignment cycle (set as 10 seconds), the MOD controller will review pending requests and if parcel delivery requests cannot be assigned to a shared ride (subject to criteria listed in SHR), and any vehicle has not been assigned any request for 1 minute or more, those vehicles are eligible to serve the parcel delivery request subject to its relative distance to the parcel origin. The maximum distance a vehicle will travel to serve a request is subject to the accepted waiting time specified for parcel delivery requests and a random vehicle is selected if there are more than one available under this threshold.
- Shared and Restricted Idle (SHR+RIDL): This is a variant of SHR+IDL aiming to minimize the impact of idle vehicle assignments on passenger rides. For this, parcel delivery services using idle vehicles (i.e., without any passenger) are limited to non-peak periods (i.e., the periods except for morning (7:00AM-10:00AM) and evening (4:00PM-9:00PM) peak). Furthermore, for passengers not to experience any stopover for parcel pickup or delivery, parcels are only picked up before the first passenger ride and dropped off after the last passenger for that ride bundle.

## 4.3 Metrics

The following *metrics* are used to evaluate relevant impacts to the mobility system, considering the perspectives of multiple agents:
- MOD travelers: total of requests served, average travel time and average waiting time.
- Shippers: total of requests served and average waiting time.
- Carriers: freight vehicle drivers' total driving time.
- MOD operator: demand served, and total distance travelled.
- Network: Vehicle km-traveled (VKT), Vehicle hour-traveled (VHT) and Travel Time Index (the ratio of simulated travel time to 'free-flow' travel time).

## 5. RESULTS

In this section we summarize the simulation results for each scenario. The displayed results are averages of outputs from three simulation replications for each scenario, to account for simulator stochasticity. First, looking at the MOD travelers in shared rides (Table 1) which are leveraged for cargo-hitching, there is a small decrease in passenger requests served (between 1% and 2%), with the higher decrease in the SHR+IDL scenario where the assignment of parcels to idle vehicles is unrestricted. Travel times, which take into account the flows of all other vehicle types (both passenger and freight vehicles) in the network, change little in the peak period (2% to 3%), but can increase up to 2 minutes in the midday period, which is expected in light of the additional freight parcel deliveries being fulfilled. This is illustrated in Figures 6 and 7, illustrating the vehicle status over time and the relation between requests and their fulfilment, respectively. Waiting times generally have small increases in the order of seconds (1% to 4%).

**Table 1 Performance metrics - MOD Shared Ride Travelers**

| Metric | Baseline | SHR | SHR+IDL | SHR+RIDL |
|---|---|---|---|---|
| **Requests served (K)** | 152 | 151 (-1%) | 149 (-2%) | 151 (-1%) |
| **Peak Travel time (min.)** | 22.5 | 22.9 (+2%) | 23.0 (+2%) | 23.3 (+3%) |
| **Midday Travel time (min.)** | 13.4 | 14.4 (+7%) | 15.0 (+10%) | 15.5 (+14%) |
| **Peak Wait time (min.)** | 5.5 | 5.7 (+4%) | 5.8 (+4%) | 5.8 (+4%) |
| **Midday Wait time (min.)** | 4.6 | 4.7 (+1%) | 4.7 (+1%) | 4.8 (+2%) |

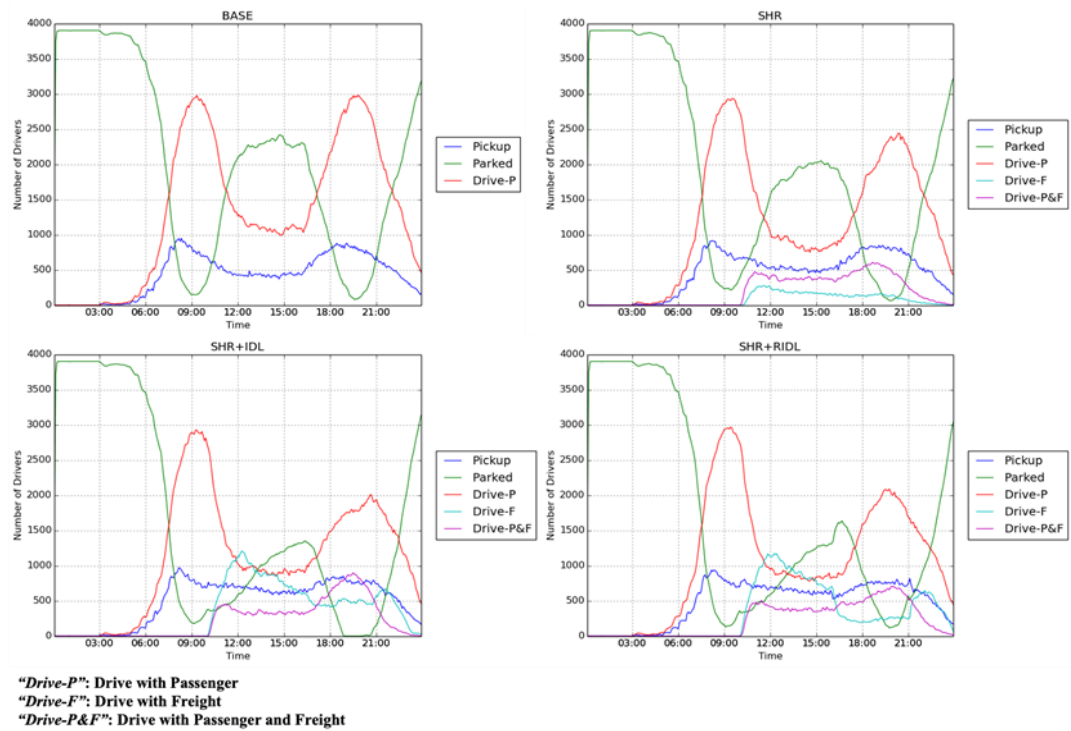

"*Drive-P*": Drive with Passenger
"*Drive-F*": Drive with Freight
"*Drive-P&F*": Drive with Passenger and Freight

**Figure 6. Plots of vehicle status over time**

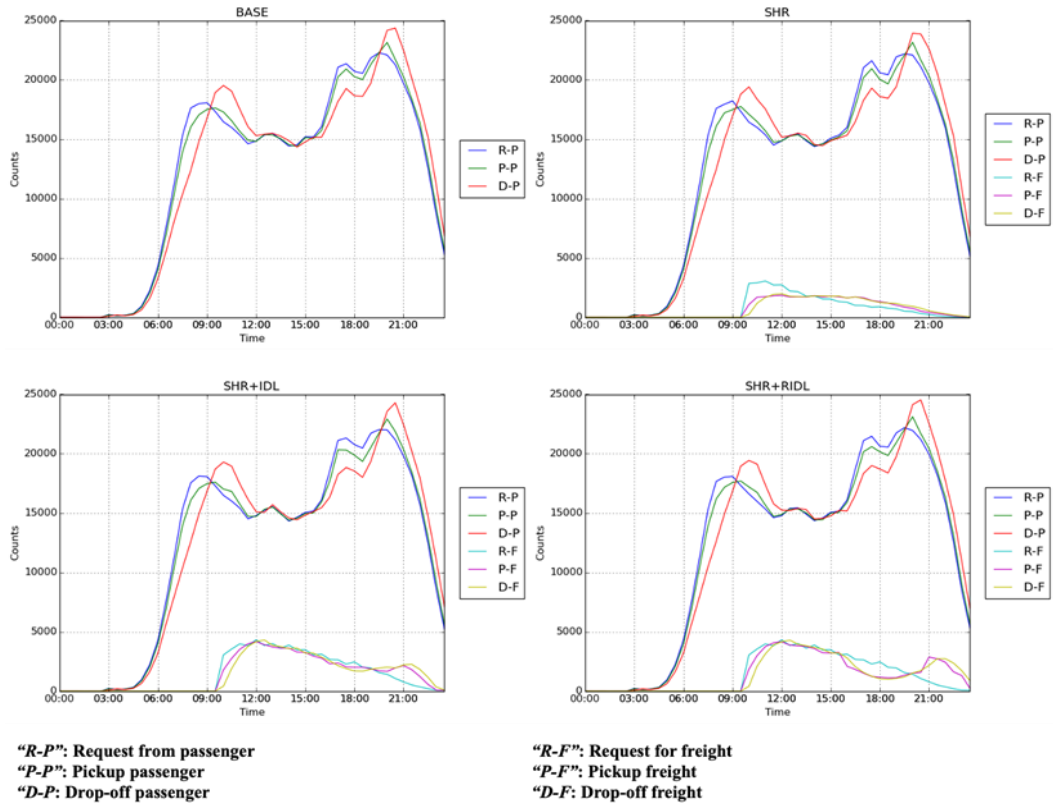

**Figure 7. Plots of requests and fulfilment over time**

The number of additional requests handled by the MOD service is shown in Table 2, which illustrates the metrics for Shippers and Carriers. The SHR scenario only allows handling a comparatively smaller number of freight deliveries, 34K instead of 67K for SHR+IDL and SHR+RIDL. Waiting times for pickups vary considerably between scenarios. Longer waiting times for pickup and delivery are important factors in acceptance of the service for both the shipper and the receiver. Larger waiting times are seen for the AM and midday periods in the SHR scenario, since a match with a shared ride needs to take place. However, in SHR+IDL and SHR+RIDL, waiting times in the AM peak and midday periods decrease substantially. In the PM peak period, for SHR+IDL, if no sharing match can be found at the time of the request, the matching algorithm defaults to looking for an idle vehicle. However, demand for passenger transport is high in the PM peak, and compared to the SHR scenario, SHR+IDL is dealing with almost twice the number of requests. Thus, on average, the time taken to serve each request increases. This can also be observed in Figure 6, where it is visible that all vehicles are busy during the PM peak, and in Figure 7, where it can be seen that as the PM peak comes to an end there is a peak in the number of freight requests being served due to a backlog. For SHR+RIDL, the peak periods are restricted for passenger trips and thus the waiting times for parcel deliveries increases even more since it does not allow using any potential vehicle that has not been assigned a request right away. Note that the passengers' waiting times remained within the reasonable boundary across the scenarios as shown in Table 1. Additionally, as expected, the driving time by the carriers decreased, between 4% to 6% due to the offloaded demand to the MOD service, meaning that the deliveries requested for the MOD service (Requests served) is now handled by the MOD operator.

**Table 2 - Performance metrics - Shippers and Carriers**

| Agent | Metric | Baseline | SHR | SHR+IDL | SHR+RIDL |
|---|---|---|---|---|---|
| Shippers | Requests served (total, K) | NA | 34 | 67 | 67 |
| | AM Peak Wait time (min.) | NA | 93.4 | 17.3 | 14.5 |
| | Midday Wait time (min.) | NA | 94.3 | 27.5 | 23.6 |
| | PM Peak Wait time (min.) | NA | 23.8 | 55.7 | 109.4 |
| Carriers | Driving time (hr, K) | 376 | 360 (-4%) | 357 (-5%) | 355 (-6%) |

*"NA" - Not Applicable; "K" - Thousands*

For the MOD operator (Table 3), results show that leveraging extra capacity of vehicles allows serving up to 11% more requests (summing passenger and freight). However, this seems to be to some extent inefficient, illustrated by the respective increase in distance travelled (17%). Still, this is not surprising as comparatively a freight vehicle tour can handle more parcels, and we assumed one parcel would take the equivalent of a passenger seat. Still, the potential for performing parcel deliveries as means to reduce the discrepancy in vehicle usage between peaks and the mid-day period is also evident. Increases in midday vehicle utilization are due to the additional demand being handled, reaching increases up to 81% from the base case.

**Table 3 - Performance metrics - MOD operator**

| | Metric | Baseline | SHR | SHR+IDL | SHR+RIDL |
|---|---|---|---|---|---|
| | Demand served (total, K) | 560 | 592 (+5%) | 624 (+10%) | 626 (+11%) |
| Shared ride | Distance travelled (km, M) | 1.9 | 2 (+5%) | 2.3 (+17%) | 2.3 (+17%) |
| | Vehicle Utilization Peak (avg.) | 77% | 80% (+4%) | 87% (+13%) | 83% (+8%) |
| | Vehicle Utilization Midday (avg.) | 42% | 53% (+26%) | 76% (+81%) | 76% (+81%) |

*"K" - Thousands; "M" - Millions*

With regards to network performance (Table 4), the shift of parcel delivery demand to MOD vehicles results in a small reduction in total VKT travelled by MOD and freight vehicles. Because of trade-off effects between the trips produced by MOD and freight vehicles, there were similar congestion levels across scenarios. The average travel time index (distance weighted TTI) is around 1.5 over 24 hours. It increases significantly during the peak periods up to 2.6 to 2.7, but the increase in VKT of MOD vehicles is counteracted by the reduction in the travelled distance by freight drivers while maintaining apparently similar MOD service levels for passengers.

**Table 4 - Performance metrics - Network**

| Users | Metric | Baseline | SHR | SHR+IDL | SHR+RIDL |
|---|---|---|---|---|---|
| MOD + Freight | Total VHT (hour, K) | 525 | 520 (-1%) | 534 (+2%) | 532 (+1%) |
| MOD + Freight | Total VKT (km, M) | 27 | 26.5 (-2%) | 26.6 (-2%) | 26.5 (-2%) |
| Overall | TTI (Weighted) Average | 1.52 | 1.51 | 1.51 | 1.50 |
| Overall | TTI (Weighted) Peak | 2.72 | 2.64 | 2.67 | 2.65 |

*"K" - Thousands; "M" - Millions*

## 5. CONCLUSIONS

In this paper we have first applied an agent-based simulation framework to systematically investigate the impacts of cargo-hitching applied to Mobility-On-Demand services. We considered the perspective of travelers, carriers, and regulators, while exploring a few illustrative assignment strategies named *shared*, *shared and idle*, *shared and restricted idle*. The results provided valuable insights about the order of magnitude of parcel movements that can be absorbed by a MOD system given a certain modal share, as well as the impacts on level of service for passengers and fleet usage. Overall, there is potential for delivering parcels using MOD vehicles with small impact to passenger travel and while reducing freight vehicle VHT and VKT, subject to operational settings. Despite a small decrease in passenger rides, there is a net increase in the number of requests handled by the MOD operator. This might be beneficial for drivers due to a more stable flow of requests across the day, which also reflects in increased usage of vehicles. However, this is the case due to the demand profiles being complementary to each other. This might or might not hold under future activity patterns of e-commerce and telework adoption. In our simulation-based analysis, the assignment methods were designed so that the impact on the passenger service level was minimal, and results indicated that those methods could fulfill this goal. However, our assumption regarding a parcel taking up a passenger seat could have contributed to this. Had we assumed more parcels could be assigned to a vehicle, this could have led to higher impacts on the passenger level of service. Still, it should be noted that the assumed fleet size is equivalent to, or smaller than, current mobility-on-demand fleets in the case-study area. Thus, our estimates are considered relatively conservative in terms of vehicle availability to perform parcel deliveries.

There are several improvements that we envision for future research. Regarding simulation realism, we ought to first enhance the B2C freight demand scenario with dedicated e-commerce demand models accounting for all B2C parcel demand, inclusive of non-same day deliveries. The assumed total of parcel demand assigned to MOD vehicles is expected to have different influences on the simulation results by operational scenario. For *shared* rides, increased demand could have higher impacts on passenger service levels. For cases using *idle* vehicles, while a higher vehicle usage could be achieved, especially in the midday period, we are unsure whether the increases in MOD VKT would be smaller than the decreases in Freight VKT. This is something we put forward to explore in the future. Regarding the analysis of operational scenarios, there is potential to explore more sophisticated settings such as having a defined time to search for sharing matches and only then defaulting to using idle vehicles, or allowing single rides to also move freight parcels as long as these are picked up before the passenger and dropped-off following its drop-off. Lastly, the application of the cargo-hitching concept to other modes such as bus and metro, or even upcoming alternatives such as automated mobility on demand could be worth exploring.

## DECLARATION

**Funding**

The research was supported in part by the National Research Foundation, Prime Minister's Office, Singapore, under its CREATE programme, Singapore-MIT Alliance for Research and Technology (SMART) Future Urban Mobility (FM) IRG. This work was also supported in part by the SUTD-MIT International Design Centre (IDC). We acknowledge the support of the Land Transport Authority (LTA) of Singapore in providing the road network for this study. Any findings, conclusions, recommendations, or opinions expressed are those of the authors only.

**Conflicts of interest/Competing interests**

On behalf of all authors, the corresponding author states that there is no conflict of interest.

**Availability of data and material**

The used data is not available for sharing due to confidentiality agreements.

**Code availability**

Part of the code used is available under an Open Source license at: https://github.com/smart-fm/simmobility-prod

**Authors' contributions**

The authors confirm contribution to the paper as follows: study conception and design:
André Romano Alho, Ravi Seshadri, Lynette Cheah, Moshe-Ben Akiva; data collection: Julia Caravias, André Romano Alho; analysis and interpretation of results: André Romano Alho, Simon Oh, Cheng Cheng, Yusuke Hara, Julia Caravias, Ravi Seshadri; software: Wen Han, Takanori Sakai, André Alho, Ravi Seshadri; draft manuscript preparation: André Romano Alho, Takanori Sakai; research funding: André Romano Alho, Lynette Cheah, Moshe Ben-Akiva. All authors reviewed the results and approved the final version of the manuscript.